\documentclass[twocolumn,english,aps,prd,reprint,floatfix,notitlepage,footinbib,preprintnumbers,superscriptaddress,longbibliography]{revtex4-1}
\pdfoutput=1
\usepackage{lmodern}

\usepackage[T1]{fontenc}
\usepackage[latin9]{inputenc}
\usepackage{geometry}
\usepackage{subfigure,lmodern, amsmath,amssymb, graphicx, pifont, adjustbox, bm, xcolor}
\usepackage{amsfonts}
\usepackage{enumitem}
\usepackage{comment}
\usepackage{mathtools}
\usepackage{float}
\usepackage{slashed}
\usepackage{ragged2e}
\usepackage{array}
\usepackage{microtype}
\usepackage{soul}
\usepackage{slashed}
\usepackage{nameref}
\usepackage{hhline}
\usepackage[unicode=true,pdfusetitle,
 bookmarks=true,bookmarksnumbered=false,bookmarksopen=false,
 breaklinks=false,pdfborder={0 0 1},backref=false,colorlinks=true]{hyperref}

\newcommand{\mysec}[1]{\noindent {\bf #1.}---}

\newcommand{\lam}{\lambda}
\newcommand{\eps}{\epsilon}
\newcommand{\Tr}{\mbox{Tr}}
\newcommand{\rank}{\mbox{rank}}
\newcommand{\diag}{\mbox{diag}}
\newcommand{\vareps}{\varepsilon}

\newcommand{\bs}[1]{\boldsymbol{#1}}

\begin{document}


\title{Helicity-Resolved Constructibility from Unified Ward Identities}

\author{Jaehoon Jeong}
\affiliation{\scalebox{1}{Cosmology, Gravity and Astroparticle Physics Group, CTPU, Institute for Basic Science, Daejeon 34126, Korea}}
\begin{abstract}
We show that the large-$z$ behavior of massive higher-spin particles under all-line transverse shifts is controlled by the helicity magnitude rather than the total spin. The apparent additional $z$ growth of nonextremal higher-spin wavefunctions is removed by unified Ward identities acting independently on each vector index. To expose this structure, we embed a four-dimensional massive momentum and its mass into an auxiliary null five-vector, whose fifth-current components characterize the breaking and restoration of the massless Ward identities. These identities determine the fixed-helicity large-$z$ scaling and yield a helicity-resolved constructibility criterion valid for arbitrary spin. Provided the stripped kinematic tensor remains finite in the massless limit, increasing the total spin does not by itself worsen constructibility, and amplitudes with nonnegative coupling dimension are constructible for all helicity configurations at five points and beyond.
\end{abstract}

\maketitle
\mysec{Introduction}
On-shell recursion reconstructs scattering amplitudes from lower-point
data when the amplitude vanishes under a suitable complex momentum
deformation at large $z$
\cite{Britto:2004ap,Britto:2005fq,Cohen:2010mi}.
The spinor-helicity formalism provides a natural language for such
deformations by making Lorentz symmetry and little-group covariance
manifest
\cite{Mangano:1990by,Dixon:1996wi,Arkani-Hamed:2017jhn}.
For massive particles, spin states are organized by the $SU(2)$
little group
\cite{Arkani-Hamed:2017jhn,Conde:2016vxs,Durieux:2020gip,
Liu:2023jbq,Jeong:2026dzt}.
Massive amplitudes are more subtle because their external
wavefunctions can depend on the momentum deformation.

The all-line transverse (ALT) shift \cite{Ema:2024vww} deforms each
external momentum along a transverse polarization vector while
preserving the on-shell conditions and momentum conservation. For
spins $s\leq1$, the shift can be chosen so that the external
wavefunctions are independent of the deformation parameter $z$
\cite{Ema:2024vww,Jeong:2026dzt}. Higher spin is more subtle.
A nonextremal-helicity state contains products of differently
polarized vectors and spinors through its Clebsch--Gordan (CG)
decomposition. Some of these components acquire additional powers of
$z$ under the ALT shift \cite{Ema:2025qgd}. Previous higher-spin
applications therefore focused on extremal-helicity states, leaving
the shifted large-$z$ behavior of nonextremal states unresolved
\cite{Ema:2025qgd}. This makes the large-$z$ behavior appear
increasingly sensitive to the total spin.

We show that this apparent spin dependence disappears. The large-$z$
behavior of a fixed-helicity state is controlled by the helicity
magnitude, not by the total spin. The mechanism is an index-by-index
Ward suppression. Each vector polarization aligned with the shift
gains one inverse power of $z$ after contraction with the current.
The CG condition then fixes a universal large-$z$ scaling for each
helicity. This resolves the nonextremal-helicity problem directly
under the ALT shift and gives a helicity-resolved constructibility
criterion for arbitrary spin, complementary to general all-line
constructibility bounds \cite{Cohen:2010mi}.

To make this structure manifest, we combine the four-dimensional
momentum and mass into the auxiliary null five-vector
$P^M=(p^\mu,im)$. Its rank-two bispinor factorization gives an
$SU(2)$ massive-spinor doublet and reduces smoothly to the two chiral
massless spinors as $m\to0$. The fifth direction is auxiliary. Its
component is fixed by the four-dimensional mass and obeys no
independent momentum-conservation law
\cite{Chiodaroli:2022ssi,Pokraka:2024fao}. This extends the auxiliary
five-vector representation used for massive-vector polarizations in
Ref.~\cite{Jeong:2025hwj}.

Reference-vector independence gives a unified Ward identity on each
vector index of the higher-spin current. The fifth-current components
measure the breaking of the corresponding massless Ward identities.
Their mass scaling determines how these identities are restored as
$m\to0$. For spin one, the construction reproduces the generalized
Ward identity familiar from Higgs and St\"uckelberg theories
\cite{Taylor:1971ff,Slavnov:1972fg,Bagger:1989fc,
Wulzer:2013mza,Cuomo:2019siq,Chen:2022gxv,Chen:2022xlg,
Kribs:2022ijz,Jeong:2025hwj}. Related Ward identities have also
appeared in constructions based on massive higher-spin gauge symmetry
\cite{Cangemi:2023XXX}. Here, however, the identity follows
kinematically from reference-vector independence and does not require
a specific mass-generation or higher-spin gauge mechanism. The same
structure extends directly to arbitrary spin.

The unified Ward identities therefore control two limits at once.
They organize the restoration of the massless Ward identities as
$m\to0$, and they remove the additional large-$z$ growth of
higher-spin wavefunctions under the ALT shift. Provided the stripped
kinematic tensor remains finite in the massless limit, increasing the
total spin does not by itself worsen constructibility. For a product
of couplings with nonnegative mass dimension, the resulting bound
guarantees constructibility for every helicity configuration at five
points and beyond.

\medskip
\mysec{Unified Spinors}We begin by embedding a four-dimensional
massive momentum into the auxiliary complex null five-vector
\begin{align}
P^M
\equiv
(p^\mu,im),
\qquad
M=0,1,2,3,4,
\end{align}
with $\eta_{MN}=\diag(1,-1,-1,-1,+1)$. The five-vector inner
product is\footnote{We use the mostly-minus signature for the four-vector space.}
\begin{align}
P_i\cdot P_j
\equiv
\eta_{MN}P_i^M P_j^N
=
p_i\cdot p_j-m_i m_j.
\label{eq:metric}
\end{align}
Thus, the four-dimensional on-shell condition $p_i^2=m_i^2$ becomes
the null condition $P_i^2=0$. The fifth direction is singled out by
\begin{align}
N^M=(0,0,0,0,1),
\quad\;
N^2=1,
\quad\;
N\cdot P=im.
\end{align}
Note that this construction does not describe genuine five-dimensional
spacetime kinematics: it retains only the four-dimensional Lorentz symmetry
characterized by $\Lambda^M{}_N N^N=N^M$, and no momentum conservation
is imposed in the fifth direction.

We represent the associated Clifford algebra by
\cite{Chiodaroli:2022ssi,Pokraka:2024fao}
\begin{gather}
(\Gamma^M)_A{}^B
\equiv
(\gamma^\mu,\gamma^5)_A{}^B,
\nonumber\\
\{\Gamma^M,\Gamma^N\}
=
2\eta^{MN}\mathbf 1_4,
\qquad
\Tr(\Gamma^M\Gamma^N)
=
4\eta^{MN}.
\end{gather}
The associated slashed momentum is
\begin{align}
\slashed P
\equiv
P_M\Gamma^M
=
\slashed p+im\gamma^5,
\qquad
\slashed P^2=0.
\label{eq:slashP}
\end{align}
Let $r=\rank(\slashed P)$. The equation $\slashed P v=0$ has
$4-r$ linearly independent solutions. Meanwhile,
$\slashed P^2=0$ implies that every column of $\slashed P$ is itself
a solution. By the definition of its rank, $r$ of these columns are
linearly independent. Therefore, we get $r\leq4-r$, giving $r\leq2$.

In the Weyl basis with
$\gamma^5=\diag(-\bs{1}_2,\bs{1}_2)$, we have
\begin{align}
\slashed P_A{}^B
&=
\begin{pmatrix}
-im\,\delta_\alpha{}^\beta
&
p_{\alpha\dot\beta}
\\[3pt]
p^{\dot\alpha\beta}
&
+im\,\delta^{\dot\alpha}{}_{\dot\beta}
\end{pmatrix}.
\label{eq:slashP_weyl}
\end{align}
For $m\neq0$, the upper-left block provides two independent rows,
giving $r\geq2$.\footnote{Although both diagonal blocks are
invertible, their ranks cannot be added because the off-diagonal
blocks relate the upper and lower rows. The on-shell relation
$p^{\dot\alpha\gamma}p_{\gamma\dot\beta}
=m^2\delta^{\dot\alpha}{}_{\dot\beta}$ gives
\begin{align*}
\bigl(
p^{\dot\alpha\beta},
im\,\delta^{\dot\alpha}{}_{\dot\beta}
\bigr)
=
\frac{i}{m}p^{\dot\alpha\gamma}
\bigl(
-im\,\delta_\gamma{}^\beta,
p_{\gamma\dot\beta}
\bigr).
\end{align*}
Thus, the lower two rows are linear combinations of the upper two.}
Together with $r\leq2$, this fixes
\begin{align}
\rank(\slashed P)
=
2.
\label{eq:slashP_rank}
\end{align}

To factorize $\slashed P$, we introduce the symplectic metric
\footnote{A symplectic metric is antisymmetric, invertible, and
invariant, $S^T\Omega S=\Omega$. The usual
$\epsilon_{\alpha\beta}$ is its two-component version; both raise
and lower spinor indices.}
\begin{align}
\Omega_{AB}
&=
\begin{pmatrix}
\eps_{\alpha\beta}&0\\
0&\eps^{\dot\alpha\dot\beta}
\end{pmatrix},
&
\Omega^{AB}
&=
\begin{pmatrix}
\eps^{\alpha\beta}&0\\
0&\eps_{\dot\alpha\dot\beta}
\end{pmatrix},
\end{align}
satisfying $\Omega^{AC}\Omega_{CB}=\delta^A{}_B$. Lowering the
second spinor index gives the antisymmetric momentum bispinor
\begin{align}
P_{AB}
\equiv
-\slashed P_A{}^C\Omega_{CB},
\qquad
P_{AB}=-P_{BA}.
\label{eq:momentum_bispinor}
\end{align}
The invertibility of $\Omega^{AB}$ gives
$\rank(P_{AB})=2$. Thus, we can define
\begin{align}
P_{AB}
&=
\eps_{IJ}|P^I)_A|P^J)_B
=
|P^I)_A|P_I)_B,
\label{eq:bispinor}
\end{align}
in terms of two independent spinors with $I=+,-$, where we choose
$\eps^{+-}=-\eps^{-+}=\eps_{-+}=-\eps_{+-}=1$.

Next, we define the symplectic dual and bracket by
\begin{align}
(P^I|^A
&\equiv
\Omega^{AB}|P^I)_B,
&
(P^I Q^J)
&\equiv
(P^I|^A|Q^J)_A.
\label{eq:symplectic_dual_bracket}
\end{align}
Then, we have
$\slashed P_A{}^B=-P_{AC}\Omega^{CB}
=|P^I)_A(P_I|^B$, and its nilpotency implies
\begin{align}
(\slashed P^2)_A{}^B
&=
|P^I)_A(P_I P^J)(P_J|^B
=
0.
\label{eq:slashP_nilpotency}
\end{align}
Here, $|P^I)_A$ and $(P_J|^B$ are rank-two $4\times2$ and
$2\times4$ matrices, respectively, so the scalar bilinear
vanishes:\footnote{The independent spinors $|P^I)$, with
$I=+,-$, allow us to construct a $2\times4$ matrix $X_I{}^A$
whose rows form a dual basis satisfying
$X_I{}^A|P^J)_A=\delta_I{}^J$. An analogous $4\times2$ matrix
exists for their symplectic duals. Multiplying
Eq.~\eqref{eq:slashP_nilpotency} by these matrices isolates
$(P_I P^J)$, which must therefore vanish.}
\begin{align}
(P^I P^J)=0.
\label{eq:scalar_bilinear}
\end{align}
Together with Eq.~\eqref{eq:bispinor}, Eq.~\eqref{eq:scalar_bilinear} gives the
Dirac equation for each unified spinor:
\begin{align}
\slashed P|P^I)=0.
\label{eq:Dirac}
\end{align}
For the vector bilinear, the antisymmetry
$(\Gamma^M)^{AB}=-(\Gamma^M)^{BA}$
\footnote{The matrices $\Gamma^M$ satisfy
$(\Gamma^M\Omega)^T=-\Gamma^M\Omega$, or equivalently,
$(\Gamma^M)_A{}^C\Omega_{CB}
=-(\Gamma^M)_B{}^C\Omega_{CA}$.}
ensures
$(P^I|\Gamma^M|P^J)=-(P^J|\Gamma^M|P^I)$,
and hence
$(P^I|\Gamma^M|P^J)\propto\eps^{IJ}$. Thus, we get
\begin{align}
(P^I|\Gamma^M|P^J)
&=
-\frac{1}{2}\eps^{IJ}\eps_{LK}
(P^K|\Gamma^M|P^L)
\nonumber\\
&=
\frac{1}{2}\eps^{IJ}\Tr(\Gamma^M\slashed P)
=
2\eps^{IJ}P^M.
\label{eq:vector_bilinear}
\end{align}

We next introduce the chiral projectors
\begin{align}
\Pi_\pm
\equiv
\frac{1\pm\slashed N}{2},
\qquad
\slashed N
\equiv
N_M\Gamma^M
=
\gamma^5,
\end{align}
which satisfy $\Pi_\pm^2=\Pi_\pm$ and
$\slashed N\Pi_\pm=\pm\Pi_\pm$. The projected spinors and brackets
are
\begin{align}
|i^I)_\pm
&\equiv
\Pi_\pm|i^I),
&
(i^I j^J)_\pm
&\equiv
(i^I|\Pi_\pm|j^J).
\label{eq:chiral_brackets}
\end{align}
The full bracket then decomposes as
$(i^I j^J)=(i^I j^J)_+ +(i^I j^J)_-$. Combining
$(P^I P^J)=0$ with
$(P^I|\slashed N|P^J)=2im\eps^{IJ}$, we have
\begin{align}
(P^I P^J)_\pm
=
\pm im\eps^{IJ}.
\label{eq:chiral_same_leg}
\end{align}
Together with Eq.~\eqref{eq:chiral_same_leg}, the Schouten identity
\footnote{In Eq.~\eqref{eq:chiral_det}, we use
$|i^I)_\pm(j^Jk^K)_\pm
+|j^J)_\pm(k^Ki^I)_\pm
+|k^K)_\pm(i^Ij^J)_\pm=0$
for the chiral brackets.}
fixes the determinant of the chiral-bracket matrix:
\begin{align}
\det\!\left[(i^I j^J)_\pm\right]
=
-m_i m_j.
\label{eq:chiral_det}
\end{align}
Thus, the chiral-bracket matrix is invertible for nonzero masses and
becomes singular in the massless limit.

In the massless limit $m\to0$, Eq.~\eqref{eq:slashP_weyl} becomes
\begin{align}
\slashed P
\to
\begin{pmatrix}
0&p_{\alpha\dot\beta}\\
p^{\dot\alpha\beta}&0
\label{eq:slashP_massless}
\end{pmatrix}.
\end{align}
Eq.~\eqref{eq:slashP} implies
$p_{\alpha\dot\alpha}p^{\dot\alpha\beta}=0$. Therefore, for a
nonzero momentum, each off-diagonal Weyl block has rank one, while
the full matrix $\slashed P$ remains rank two. The two independent
spinors in Eq.~\eqref{eq:bispinor} may then be chosen as
\begin{align}
|i^-)_-
&\to
\begin{pmatrix}
|i\rangle\\
0
\end{pmatrix},
&
|i^+)_+
&\to
\begin{pmatrix}
0\\
|i]
\end{pmatrix}.
\label{eq:massless_spinors}
\end{align}
Thus, the unified doublet reduces to two independent chiral massless
spinors. Substituting Eq.~\eqref{eq:massless_spinors} into
Eq.~\eqref{eq:bispinor}, Eq.~\eqref{eq:slashP_massless} gives
\begin{align}
p_{\alpha\dot\alpha}
=
|p\rangle_\alpha[p|_{\dot\alpha},
\end{align}
which is the conventional massless spinor-helicity
factorization~\cite{Dixon:1996wi}.

\medskip
\mysec{Unified Ward Identities}
We define the spin-1 polarization vector in terms of a reference
vector $R^M$ with $P\cdot R\neq0$:
\begin{align}
\varepsilon^{IJ,M}(P,R)
&\equiv
\frac{1}{2P\cdot R}
(P^{(I}|\Gamma^M\slashed R|P^{J)}).
\label{eq:unified_vector_polarization}
\end{align}
Here, parentheses denote symmetrization with weight $1/2$.
The relations $\slashed P|P^I)=0$, $(P^I P^J)=0$, and
$\slashed R^2=R^2\mathbf1$ give
\begin{align}
P\cdot \varepsilon^{IJ}(P,R)
&=
R\cdot \varepsilon^{IJ}(P,R)
=
0,
\label{eq:unified_polarization_transversality}
\end{align}
without requiring $R^2=0$.

The Dirac equation and the Clifford algebra identities
$\{\slashed P,\slashed R\}=2P\cdot R$ and
$\{\slashed P,\slashed R'\}=2P\cdot R'$ also imply
\begin{align}
&(P\cdot R)
(P^{(I}|\Gamma^M\slashed R'|P^{J)})
-
(P\cdot R')
(P^{(I}|\Gamma^M\slashed R|P^{J)})
\nonumber\\
&=
-P^M
(P^{(I}|\slashed R'\slashed R|P^{J)}).
\label{eq:reference_spinor_identity}
\end{align}
Dividing this identity by $2(P\cdot R)(P\cdot R')$, we get
\begin{align}
\varepsilon^{IJ}(P,R')
&=
\varepsilon^{IJ}(P,R)
+
\zeta^{IJ}(P;R',R)P,
\label{eq:reference_shift}
\end{align}
where
\begin{align}
\zeta^{IJ}(P;R',R)
&=
-\frac{
R'\cdot\varepsilon^{IJ}(P,R)
}{
P\cdot R'
}.
\end{align}
Thus, changing the reference vector shifts every symmetric
little-group component along $P^M$, including the longitudinal
component $(I,J)=(+,-)$.

For a simple analysis, we define the helicity polarizations as
\begin{align}
\left(
\varepsilon^{+},
\varepsilon^{0},
\varepsilon^{-}
\right)
&\equiv
\left(
\varepsilon^{++},
\sqrt{2}\,\varepsilon^{+-},
\varepsilon^{--}
\right).
\label{eq:vector_helicity_polarization}
\end{align}
Next, we denote the current before contraction with the external
polarization by $J_M$. The component $J_4$
completes the five-vector current for a general reference vector.
It transforms as a four-dimensional Lorentz scalar but does not
represent an additional physical scalar state.

Under $R\to R'$, the amplitude
$A^\lam=\varepsilon^{\lam,M}J_M$ changes by
\begin{align}
\delta A^\lam
&=
\zeta^\lam P^M J_M.
\end{align}
The reference vector does not belong to the physical kinematics, so
the amplitude cannot depend on its choice. Thus, reference-vector
independence requires $\delta A^\lam=0$ and gives the unified Ward
identity
\begin{align}
P^M J_M
&=
0,
\qquad
p^\mu J_\mu=-imJ_4.
\label{eq:unified_Ward}
\end{align}
The second relation makes the role of the fifth-current component
particularly transparent: {\it $mJ_4$ measures the breaking of the four-dimensional massless Ward identity caused by the mass deformation.} 

The massless Ward identity is recovered whenever
\begin{align}
mJ_4\to0,
\label{eq:spin1_massless_Ward}
\end{align}
for $m\to0$. Thus, the mass scaling of $J_4$ determines how the
massless Ward identity is restored: a finite $J_4$, or more generally
a divergence weaker than $1/m$, preserves the Ward identity, whereas a $1/m$ or stronger divergence obstructs this limit.

Note that $J_4$ is required by reference-vector independence,
rather than introduced by hand through $p^\mu J_\mu$. It transforms
as a four-dimensional Lorentz scalar, while its physical interpretation
is theory-dependent. For perturbative local theories with meromorphic
mass dependence, the condition $mJ_4\to0$ further implies regularity
of $J_4$ at $m=0$, as discussed in the Appendix.

We next turn to the unified Ward identities for higher spins. For a bosonic particle, we combine $s$ vector polarizations into
the spin-$s$, helicity-$h$ wavefunction:
\begin{align}
\varepsilon^{h}_{M_1\cdots M_s}(P,\{R\})
&=
\sum_{\{\lam_a\}}
\left\langle
1\lam_1,\ldots,1\lam_s
\middle|
sh
\right\rangle
\nonumber\\
&\quad\times
\varepsilon^{\lam_1}_{M_1}(P,R_1)
\cdots
\varepsilon^{\lam_s}_{M_s}(P,R_s),
\label{eq:integer_spin_helicity_polarization}
\end{align}
where $\lam_a=0,\pm1$ $(a=1,\ldots,s)$ and
$\{R\}=(R_1,\ldots,R_s)$. The Clebsch--Gordan (CG) coefficient
projects onto spin $s$ and helicity $h$ and requires
$\sum_a\lam_a=h$. For a fermion of spin
$s=\ell+\tfrac12$, we couple the spin-$\ell$ polarization in
Eq.~\eqref{eq:integer_spin_helicity_polarization} to a unified
spinor:
\begin{align}
&u^{h}_{M_1\cdots M_\ell}(P,\{R\})
\nonumber\\
&=
\sum_{\lam=-\ell}^{\ell}
\sum_{\kappa=\pm}
\left\langle
\ell\lam,\tfrac12\kappa
\middle|
sh
\right\rangle
\varepsilon^{\lam}_{M_1\cdots M_\ell}(P,\{R\})
|P^\kappa),
\label{eq:half_integer_spin_helicity_polarization}
\end{align}
where $\kappa=\pm$ is identified with $\pm1$, and the CG coefficient
requires $\lam+\kappa/2=h$.

For higher spins, the CG coefficients construct wavefunctions that
are symmetric under simultaneous permutations of $(M_a,R_a)$ and
traceless in the five-vector indices, even for independent reference
vectors. Indeed, reference shifts preserve tracelessness because they
are aligned with $P^M$ and satisfy
$P\cdot\varepsilon^\lam=P^2=0$. Thus, we take the tensor currents
$J_{M_1\cdots M_\ell}$ coupled to these wavefunctions to be symmetric
and traceless.

Using reference-vector independence for each polarization, we have
\begin{align}
P^{M_r}
J_{M_1\cdots M_r\cdots M_\ell}
&=
0,
\qquad
r=1,\ldots,\ell,
\label{eq:all_index_unified_Ward}
\end{align}
where $\ell=s$ and $s-\tfrac12$ for bosons and fermions,
respectively. Here, we omit the other polarizations contracted with
$J_{\{M\}}$, together with the spinor for fermions. Varying multiple
reference vectors, we get identities with multiple momentum
contractions.

To display the four-dimensional Ward chain, we write
\begin{align}
J_{4^r\mu_{r+1}\cdots\mu_\ell}
&\equiv
J_{\underbrace{\raisebox{-0.2ex}{$\scriptstyle 4\cdots4$}}_{r}
\mu_{r+1}\cdots\mu_\ell}.
\end{align}
Each momentum contraction replaces one vector index by a fifth
index and supplies a factor of $-im$:
\begin{align}
&p^{\mu_1}\cdots p^{\mu_r}
J_{\mu_1\cdots\mu_\ell}
=
(-im)p^{\mu_2}\cdots p^{\mu_r}
J_{4\mu_2\cdots\mu_\ell}
\nonumber\\
&=
\cdots
=
(-im)^r
J_{4^r\mu_{r+1}\cdots\mu_\ell},
\qquad
1\leq r\leq\ell.
\label{eq:iterated_massive_Ward}
\end{align}
Under the mass scaling $m=\tau\bar m$, the Ward chain becomes
\begin{align}
p^{\mu_r}
J_{4^{r-1}\mu_r\cdots\mu_\ell}
&=
-i\tau\bar m
J_{4^r\mu_{r+1}\cdots\mu_\ell}.
\label{eq:scaled_Ward}
\end{align}
If $J_{4^r\mu_{r+1}\cdots\mu_\ell}$ remains finite as
$\tau\to0$, the right-hand side vanishes. Thus, in this limit, we have
\begin{align}
p^{\mu_r}
J_{4^{r-1}\mu_r\cdots\mu_\ell}
\;\;\to\;\;
0.
\label{eq:higher_spin_massless_Ward}
\end{align}
For $r=1$, this gives the usual massless Ward identity for every
vector index of the higher-spin current. 

For $r>1$, Eq.~\eqref{eq:higher_spin_massless_Ward} gives a sequence
of Ward identities for current components carrying fifth indices.
If a current component first becomes regular at a specific $r$, a
massless Ward identity emerges at that point, while singular mass
dependence can obstruct the preceding steps of the chain. Thus, the
Ward chain identifies the first level at which a massless Ward
identity emerges within the higher-spin current.

\medskip
\mysec{Helicity-Resolved Constructibility}
We apply the ALT shift
\cite{Ema:2024vww,Ema:2025qgd} directly to the unified spinors.
Setting $\sigma_i=+$ for $h_i\geq0$ and $\sigma_i=-$ for $h_i<0$,
\footnote{Either choice can be used for $h_i=0$.}
we take
\begin{align}
|\hat i^I)
&=
|i^I)
+
2izc_i
\begin{cases}
\delta^I{}_-|i^+)_-/m_i,
& \sigma_i=+,
\\[2pt]
\delta^I{}_+|i^-)_+/m_i,
& \sigma_i=-.
\end{cases}
\label{eq:ALT_spinor_shift}
\end{align}
The ratios in Eq.~\eqref{eq:ALT_spinor_shift} remain finite in the
massless limit.\footnote{Eqs.~\eqref{eq:chiral_same_leg} and
\eqref{eq:massless_spinors} show that $|i^+)_-$ and $|i^-)_+$ vanish
linearly with $m_i$, so the ratios in
Eq.~\eqref{eq:ALT_spinor_shift} are defined by their smooth
$m_i\to0$ limits.} For each leg, we choose the fixed null reference five-vector
\begin{align}
R_{{\rm eq},i}^M
&=
(1,-\vec p/|\vec p|,0).
\label{eq:ALT_equivalent_reference}
\end{align}
This choice leaves the transverse polarizations in their conventional form, while yielding the equivalent-gauge longitudinal polarization~\cite{Wulzer:2013mza}.

Thus, the spinor deformation shifts the momentum linearly in $z$:
\begin{gather}
\hat P_{i,AB}(z)
\equiv
\eps_{IJ}
|\hat i^I)_A|\hat i^J)_B
=
P_{i,AB}
+
zQ_{i,AB},
\nonumber\\
Q_i^M
=
c_i\vareps_i^{\sigma_i,M},
\quad\;
\vareps_i^{\sigma_i,M}
\equiv
\vareps^{\sigma_i,M}(P_i,R_{{\rm eq},i}).
\label{eq:ALT_momentum_shift}
\end{gather}
For this reference choice, we get
$Q_i^2=P_i\cdot Q_i=R_{{\rm eq},i}\cdot Q_i=N\cdot Q_i=0$, and hence
\begin{gather}
\hat P_i^2(z)
=
0,
\qquad
N\cdot\hat P_i(z)
=
im_i,
\label{eq:ALT_on_shell_relations}
\end{gather}
preserving the on-shell condition. We choose the coefficients $c_i$ to satisfy
\begin{align}
\sum_{i=1}^nQ_i
&=
\sum_{i=1}^n
c_i\vareps_i^{\sigma_i}
=
0.
\label{eq:ALT_momentum_conservation}
\end{align}
The shift therefore preserves four-dimensional momentum conservation
and leaves the fifth components unchanged:
\begin{align}
\sum_{i=1}^n\hat P_i^M
&=
iM_{\rm tot}N^M,
\quad\;
M_{\rm tot}
\equiv
\sum_{i=1}^nm_i.
\end{align}

For higher spins, the CG decomposition of a fixed-helicity
wavefunction contains vector polarizations and spinors both aligned
and nonaligned with $\sigma_i$. To display their shifts, we focus on
one external leg and suppress its label. Then, 
Eq.~\eqref{eq:ALT_spinor_shift} gives
\begin{align}
|\hat P^\sigma)
&=
|P^\sigma),
\nonumber\\
|\hat P^{-\sigma})
&=
|P^{-\sigma})
+
\frac{2izc}{m}
|P^\sigma)_{-\sigma},
\label{eq:ALT_spinor_wavefunction_shift}
\end{align}
for unified spinors and
\begin{align}
\hat{\vareps}^{\sigma}
&=
\vareps^{\sigma},
\qquad
\hat{\vareps}^{0}
=
\vareps^{0},
\nonumber\\
\hat{\vareps}^{-\sigma}
&=
\vareps^{-\sigma}
+
\frac{zc}{P\cdot R_{\rm eq}}\,R_{\rm eq}.
\label{eq:ALT_vector_wavefunction_shift}
\end{align}
for the vector polarizations where we used
$R_{\rm eq}\cdot\vareps^\sigma=0$, so that
$\hat P\cdot R_{\rm eq}=P\cdot R_{\rm eq}$. In Eqs.~\eqref{eq:ALT_vector_wavefunction_shift} and~\eqref{eq:ALT_spinor_wavefunction_shift}, the $z$ dependence of unaligned spinor and vector polarization can aggravate the constructibility of
higher-spin amplitudes.

However, the unified Ward identities suppress the aligned polarization
$\vareps^\sigma$. Consider a tensor current $J_{\{M\}}$ contracted with
a higher-spin wavefunction. Using $\vareps^\sigma=(\hat P-P)/(zc)$ from Eq.~\eqref{eq:ALT_momentum_shift} and
$\hat P^{M_a}\hat J_{\cdots M_a\cdots}=0$ from
Eq.~\eqref{eq:all_index_unified_Ward}, we get
\begin{align}
\vareps^{\sigma,M_a}
\hat J_{\cdots M_a\cdots}
&=
\frac{1}{zc}
\left(
\hat P^{M_a}-P^{M_a}
\right)
\hat J_{\cdots M_a\cdots}
\nonumber\\
&=
-\frac{1}{zc}
P^{M_a}
\hat J_{\cdots M_a\cdots},
\label{eq:ALT_single_Ward_suppression}
\end{align}
where we omit the other contractions. Thus, if the tensor
current scales as $\hat J_{\{M\}}\sim z^\omega$, the higher-spin
Ward identities improve the leading large-$z$ weight by one power
for each aligned vector polarization:
\begin{align}
&
\vareps^{\sigma,M_1}\cdots
\vareps^{\sigma,M_r}
\hat J_{M_1\cdots M_r\cdots}
\nonumber\\
&=
\left(-\frac{1}{zc}\right)^r
P^{M_1}\cdots P^{M_r}
\hat J_{M_1\cdots M_r\cdots}
\sim
z^{\omega-r},
\label{eq:ALT_multiple_Ward_suppression}
\end{align}
where $r$ denotes the number of aligned vector polarizations in a
specific vector-polarization configuration.

This Ward suppression allows us to determine the large-$z$ weights of the spinors and vector polarizations after contraction with the tensor current:
\begin{gather}
|P^\sigma):\ z^0,
\qquad
|P^{-\sigma}):\ z,
\nonumber\\
\vareps^\sigma:\ z^{-1},
\qquad
\vareps^0:\ z^0,
\qquad
\vareps^{-\sigma}:\ z.
\label{eq:ALT_net_wavefunction_weights}
\end{gather}
Each pair of $\vareps^\sigma$ and $\vareps^{-\sigma}$ has zero net
weight, as does every longitudinal polarization.

The weight of a vector polarization $\vareps^\lam$ can be written
as $z^{-\sigma\lam}$. For a bosonic wavefunction, the CG condition
$\sum_a\lam_a=h$ therefore gives
$z^{-\sigma h}=z^{-|h|}$. For a fermionic wavefunction, the spinor
has weight $z^{(1-\sigma\kappa)/2}$. Using
$\sum_a\lam_a+\kappa/2=h$, the total weight becomes
\begin{align}
z^{-\sigma(h-\kappa/2)}
z^{(1-\sigma\kappa)/2}
&=
z^{-|h|+\frac12}.
\end{align}
Thus, for an $n$-point amplitude, all external wavefunctions
together contribute the large-$z$ weight
\begin{align}
z^{-\sum_{i=1}^n|h_i|+\frac{N_F}{2}},
\label{eq:external_wavefunction_scaling}
\end{align}
where $N_F$ denotes the number of external fermions.

Consider an $n$-point amplitude with a product of couplings $g$.
After stripping all external wavefunctions, we write the tensor
current as
\begin{align}
J_{\{M\}}
&=
gF_{\{M\}},
\end{align}
where $g$ is the product of the coupling constants appearing in the
interaction vertices, and $F_{\{M\}}$ is the stripped kinematic tensor, including the
propagators. Since $g$ is
independent of $z$, $J_{\{M\}}$ and $F_{\{M\}}$ have the same
large-$z$ weight. The mass dimension of the kinematic tensor is
\begin{align}
[F_{\{M\}}]
&=
4-n-[g]-\frac{N_F}{2}.
\end{align}

Assuming that the stripped kinematic tensor $F_{\{M\}}$ remains
finite in the massless limit, dimensional analysis gives
\begin{gather}
\hat F_{\{M\}}(z)
\sim
z^\omega,
\qquad
\omega
\leq
4-n-[g]-\frac{N_F}{2}.
\label{eq:dimensional_large_z_scaling}
\end{gather}
Combining this result with
Eq.~\eqref{eq:external_wavefunction_scaling}, we get
\begin{align}
\hat A_n^{h_1\cdots h_n}(z)
&\sim
z^\gamma,
\qquad
\gamma
\leq
4-n-[g]-\sum_{i=1}^n|h_i|.
\label{eq:helicity_resolved_bound}
\end{align}
Thus, the amplitude is constructible whenever
\begin{align}
4-n-[g]-\sum_{i=1}^n|h_i|
&<
0,
\end{align}
as $\hat A_n(z)$ vanishes at large $z$ and gives no boundary
contribution.

Under the condition in
Eq.~\eqref{eq:dimensional_large_z_scaling}, the bound guarantees
constructibility for every helicity configuration when $[g]\geq0$
and $n\geq5$, while leaving undetermined only the case with $n=4$,
$[g]=0$, and $h_i=0$ for every external leg. For minimal electromagnetic and gravitational Compton amplitudes,
the interactions fix the mass dimension $[g]$ through their momentum
insertions. Substituting these dimensions into
Eq.~\eqref{eq:helicity_resolved_bound}, one can reproduce the known
constructibility ranges $s\leq3/2$ and $s\leq5/2$, respectively
\cite{Ema:2025qgd,Chiodaroli:2021eug}.

\medskip
\mysec{Discussion}
We have shown that the large-$z$ behavior of a fixed-helicity higher-spin state is controlled by the helicity magnitude $|h|$, rather than the total spin $s$. Increasing the spin therefore does not by itself worsen constructibility. Together with dimensional analysis, this gives a helicity-resolved constructibility bound valid for arbitrary spin. If the stripped kinematic tensor remains finite as $m\to0$, the bound guarantees constructibility for every helicity configuration when $n\geq5$ and $[g]\geq0$.

The unified Ward identities also organize the massless limit. The fifth-current components measure the breaking of the massless Ward identities caused by the mass deformation, and their mass scaling determines how these identities are restored as $m\to0$. For higher spin, this structure extends index by index through the fifth-current hierarchy. The same Ward structure therefore controls both the approach to the massless limit and the large-$z$ behavior relevant for constructibility. An important open question is how dynamics constrains the mass scaling of the higher-spin fifth-current components.

\medskip
\noindent {\it Acknowledgments:}
We thank Alex Pomarol, Wenqi Ke, and Yu-Hui Zheng for valuable comments. J.J. is supported by IBS under the project code IBS-R018-D3.

\bibliographystyle{utphys-modified}
\bibliography{unified_spinor}

@article{Mangano:1990by,
    author = "Mangano, Michelangelo L. and Parke, Stephen J.",
    title = "{Multiparton amplitudes in gauge theories}",
    eprint = "hep-th/0509223",
    archivePrefix = "arXiv",
    primaryClass = "hep-th",
    journal = "Phys. Rept.",
    volume = "200",
    pages = "301--367",
    year = "1991"
}

@article{Dixon:1996wi,
    author = "Dixon, Lance J.",
    title = "{Calculating scattering amplitudes efficiently}",
    eprint = "hep-ph/9601359",
    archivePrefix = "arXiv",
    primaryClass = "hep-ph",
    year = "1996"
}

@article{Britto:2004ap,
    author = "Britto, Ruth and Cachazo, Freddy and Feng, Bo",
    title = "{New recursion relations for tree amplitudes of gluons}",
    eprint = "hep-th/0412308",
    archivePrefix = "arXiv",
    primaryClass = "hep-th",
    journal = "Nucl. Phys. B",
    volume = "715",
    pages = "499--522",
    year = "2005"
}

@article{Britto:2005fq,
    author = "Britto, Ruth and Cachazo, Freddy and Feng, Bo and Witten, Edward",
    title = "{Direct proof of the tree-level scattering amplitude recursion relation in Yang-Mills theory}",
    eprint = "hep-th/0501052",
    archivePrefix = "arXiv",
    primaryClass = "hep-th",
    journal = "Phys. Rev. Lett.",
    volume = "94",
    pages = "181602",
    year = "2005"
}

@article{Arkani-Hamed:2017jhn,
    author = "Arkani-Hamed, Nima and Huang, Tzu-Chen and Huang, Yu-tin",
    title = "{Scattering amplitudes for all masses and spins}",
    eprint = "1709.04891",
    archivePrefix = "arXiv",
    journal = "JHEP",
    volume = "11",
    pages = "070",
    year = "2021"
}

@article{Conde:2016vxs,
    author = "Conde, Eduardo and Marzolla, Andrea",
    title = "{Lorentz constraints on massive three-point amplitudes}",
    eprint = "1601.08113",
    archivePrefix = "arXiv",
    journal = "JHEP",
    volume = "09",
    pages = "041",
    year = "2016"
}

@article{Durieux:2020gip,
    author = "Durieux, Gauthier and Kitahara, Teppei and Machado, Camila S. and Shadmi, Yael and Weiss, Yaniv",
    title = "{Constructing massive on-shell contact terms}",
    eprint = "2008.09652",
    archivePrefix = "arXiv",
    journal = "JHEP",
    volume = "12",
    pages = "175",
    year = "2020"
}

@article{Liu:2023jbq,
    author = "Liu, Hongkai and Ma, Teng and Shadmi, Yael and Waterbury, Michael",
    title = "{An EFT hunter's guide to two-to-two scattering: HEFT and SMEFT on-shell amplitudes}",
    eprint = "2301.11349",
    archivePrefix = "arXiv",
    journal = "JHEP",
    volume = "05",
    pages = "241",
    year = "2023"
}

@article{Jeong:2026dzt,
    author = "Jeong, Jaehoon and Ko, Pyungwon and Zheng, Yu-Hui",
    title = "{Explicit Conditions for Diagnosing Tree-Level Unitarity}",
    eprint = "2605.12057",
    archivePrefix = "arXiv",
    year = "2026"
}

@article{Chiodaroli:2022ssi,
    author = "Chiodaroli, Marco and Gunaydin, Murat and Johansson, Henrik and Roiban, Radu",
    title = "{Spinor-helicity formalism for massive and massless amplitudes in five dimensions}",
    eprint = "2202.08257",
    archivePrefix = "arXiv",
    journal = "JHEP",
    volume = "02",
    pages = "040",
    year = "2023"
}

@article{Pokraka:2024fao,
    author = "Pokraka, Andrzej and Rajan, Smita and Ren, Lecheng and Volovich, Anastasia and Zhao, W. Wayne",
    title = "{Five-dimensional spinor helicity for all masses and spins}",
    eprint = "2405.09533",
    archivePrefix = "arXiv",
    journal = "JHEP",
    volume = "07",
    pages = "056",
    year = "2025"
}

@article{Taylor:1971ff,
    author = "Taylor, J. C.",
    title = "{Ward identities and charge renormalization of the Yang-Mills field}",
    journal = "Nucl. Phys. B",
    volume = "33",
    pages = "436--444",
    year = "1971"
}

@article{Slavnov:1972fg,
    author = "Slavnov, A. A.",
    title = "{Ward identities in gauge theories}",
    journal = "Theor. Math. Phys.",
    volume = "10",
    pages = "99--107",
    year = "1972"
}

@article{Bagger:1989fc,
    author = "Bagger, Jonathan and Schmidt, Carl",
    title = "{Equivalence theorem redux}",
    journal = "Phys. Rev. D",
    volume = "41",
    pages = "264",
    year = "1990"
}

@article{Wulzer:2013mza,
    author = "Wulzer, Andrea",
    title = "{An equivalent gauge and the equivalence theorem}",
    eprint = "1309.6055",
    archivePrefix = "arXiv",
    journal = "Nucl. Phys. B",
    volume = "885",
    pages = "97--126",
    year = "2014"
}

@article{Cuomo:2019siq,
    author = "Cuomo, Gabriel and Vecchi, Luca and Wulzer, Andrea",
    title = "{Goldstone equivalence and high energy electroweak physics}",
    eprint = "1911.12366",
    archivePrefix = "arXiv",
    journal = "SciPost Phys.",
    volume = "8",
    pages = "078",
    year = "2020"
}

@article{Kribs:2022ijz,
    author = "Kribs, Graham D. and Lee, Gabriel and Martin, Adam",
    title = "{Effective field theory of Stueckelberg vector bosons}",
    eprint = "2204.01755",
    archivePrefix = "arXiv",
    journal = "Phys. Rev. D",
    volume = "106",
    pages = "055020",
    year = "2022"
}

@article{Jeong:2025hwj,
    author = "Jeong, Jaehoon",
    title = "{No gauge cancellation at high energy in the five-vector $R_\xi$ gauge}",
    eprint = "2502.13633",
    archivePrefix = "arXiv",
    journal = "Phys. Rev. D",
    volume = "111",
    pages = "076031",
    year = "2025"
}

@article{Ema:2024vww,
    author = "Ema, Yohei and Gao, Ting and Ke, Wenqi and Liu, Zhen and Lyu, Kun-Feng and Mahbub, Ishmam",
    title = "{Momentum shift and on-shell constructible massive amplitudes}",
    eprint = "2403.15538",
    archivePrefix = "arXiv",
    journal = "Phys. Rev. D",
    volume = "110",
    pages = "105003",
    year = "2024"
}

@article{Ema:2025qgd,
    author = "Ema, Yohei and Gao, Ting and Ke, Wenqi and Liu, Zhen and Mahbub, Ishmam",
    title = "{On-shell recursion relations for higher-spin Compton amplitudes}",
    eprint = "2506.02106",
    archivePrefix = "arXiv",
    journal = "JHEP",
    volume = "01",
    pages = "069",
    year = "2026"
}

@article{Chiodaroli:2021eug,
    author = "Chiodaroli, Marco and Johansson, Henrik and Pichini, Paolo",
    title = "{Compton black-hole scattering for $s\leq 5/2$}",
    eprint = "2107.14779",
    archivePrefix = "arXiv",
    journal = "JHEP",
    volume = "02",
    pages = "156",
    year = "2022"
}

@article{Chen:2022gxv,
    author = "Chen, Junmou and Hagiwara, Kaoru and Kanzaki, Junichi and Mawatari, Kentarou",
    title = "{Helicity amplitudes without gauge cancellation for electroweak processes}",
    eprint = "2203.10440",
    archivePrefix = "arXiv",
    primaryClass = "hep-ph",
    reportNumber = "KEK-TH-2403, IPMU22-0008",
    doi = "10.1140/epjc/s10052-023-12093-7",
    journal = "Eur. Phys. J. C",
    volume = "83",
    number = "10",
    pages = "922",
    year = "2023",
    note = "[Erratum: Eur.Phys.J.C 84, 97 (2024)]"
}

@article{Chen:2022xlg,
    author = "Chen, Junmou and Hagiwara, Kaoru and Kanzaki, Junichi and Mawatari, Kentarou and Zheng, Ya-Juan",
    title = "{Helicity amplitudes in light-cone and Feynman-diagram gauges}",
    eprint = "2211.14562",
    archivePrefix = "arXiv",
    primaryClass = "hep-ph",
    reportNumber = "KEK-TH-2471, IPMU22-0055",
    doi = "10.1140/epjp/s13360-024-05067-5",
    journal = "Eur. Phys. J. Plus",
    volume = "139",
    number = "4",
    pages = "332",
    year = "2024"
}

@article{Cohen:2010mi,
    author = "Cohen, Timothy and Elvang, Henriette and Kiermaier, Michael",
    title = "{On-shell constructibility of tree amplitudes in general field theories}",
    eprint = "1010.0257",
    archivePrefix = "arXiv",
    primaryClass = "hep-th",
    reportNumber = "MIT-CTP-4189",
    doi = "10.1007/JHEP04(2011)053",
    journal = "JHEP",
    volume = "04",
    pages = "053",
    year = "2011"
}

@article{Cangemi:2023XXX,
    author = "Cangemi, Lucile and Chiodaroli, Marco and Johansson, Henrik
              and Ochirov, Alexander and Pichini, Paolo and Skvortsov, Evgeny",
    title = "{From higher-spin gauge interactions to Compton amplitudes for root-Kerr}",
    eprint = "2311.14668",
    archivePrefix = "arXiv",
    primaryClass = "hep-th",
    journal = "JHEP",
    volume = "09",
    pages = "196",
    year = "2024",
    doi = "10.1007/JHEP09(2024)196"
}

\medskip
\noindent {\bf Appendix.--}
The fifth-current components introduced in the main text do not in
general correspond to additional physical lower-spin states. Their
role is fixed by the unified Ward identity,
\begin{equation}
p^\mu J_\mu=-imJ_4.
\end{equation}
The ordinary massless Ward identity is recovered whenever $mJ_4\to0$ as $m\to0$. Thus, $J_4$ need not remain finite: any divergence weaker
than $1/m$ is compatible with the restoration of the massless Ward identity, while a $1/m$ or stronger divergence obstructs the identity.

For perturbative local theories with meromorphic mass dependence, however, the condition $mJ_4\to0$ implies that $J_4$ is regular at
$m=0$, since any pole would be at least of order $1/m$. A finite
$J_4$ can include the familiar Higgs and St\"uckelberg realizations, where $J_4$ is identified with the Goldstone scalar current. 

For higher spin, regularity of the fifth-index current
gives
\begin{equation}
p^{\mu_1}
J_{\mu_1\mu_2\cdots\mu_\ell}
=
O(m),
\label{eq:compton_current_constraint}
\end{equation}
which is the current constraint used in higher-spin Compton
constructions \cite{Ema:2025qgd,Chiodaroli:2021eug}.
This condition constrains only the contraction with $p^{\mu_1}$, not
the full current. In particular, the current may contain a singular
transverse contribution
\begin{equation}
J_{\mu_1\cdots\mu_\ell}
\supset
\frac{1}{m^q}
J^\perp_{\mu_1\cdots\mu_\ell},
\quad
p^{\mu_1}J^\perp_{\mu_1\cdots\mu_\ell}=0,
\quad
q>0,
\end{equation}
which is invisible to Eq.~\eqref{eq:compton_current_constraint}.
Thus, the massless Ward identity can emerge even when the full current
is singular in the massless limit. This separates the regularity of the
fifth-index current from the existence of a smooth interacting massless
higher-spin limit.

\end{document}